\documentclass{aa}
\usepackage{txfonts}
\usepackage{natbib}
\usepackage{graphicx}
\usepackage{amssymb}
\usepackage{mathrsfs}
\usepackage[dvips]{rotating}
\bibpunct{(}{)}{;}{a}{}{,}

\def\pl{\mathrm{planet}}
\def\jup{\mathrm{Jupiter}}
\def\res{$\lambda/\Delta\lambda$~}
\def\vmax{$v_\mathrm{max}$}

\def\OI{[O\,{\sc i}] }

\def\kms{$\mathrm{km}~\mathrm{s}^{-1}$}

\def\deg{$^\circ$}
\begin{document}
\title{Resolving the disk rotation of HD~97048 and HD~100546 in the
 [O\,{\sc i}] 6300\r{A} line: evidence for a giant planet orbiting
 HD~100546 \thanks{Based on observations collected at the European
 Southern Observatory, Paranal, Chile (program number 075.C-0172).}
} 
\titlerunning{Spectrally and spatially resolving the \OI line}
\authorrunning{Acke \& van~den~Ancker}
\author{B.~Acke\inst{1,}\thanks{Postdoctoral researcher of the Research
  Fund KULeuven.} \and M.~E.~van~den~Ancker\inst{2}}
\institute{Instituut voor Sterrenkunde, KULeuven, Celestijnenlaan 200B, 
3001 Leuven, Belgium\\ 
\email{Bram.Acke@ster.kuleuven.be}
\and
European Southern Observatory, Karl-Schwarzschild Strasse 2, D-85748
Garching bei M\"unchen, Germany}
\date{DRAFT, \today}

\abstract{}
{We intend to spatially and spectrally resolve the \OI emission region
in two nearby Herbig stars.}
{We present high-resolution (\res = 80,000) VLT/UVES
 echelle spectra of the \OI 6300\r{A} line in the Herbig Ae/Be stars
 \object{HD~97048} and \object{HD~100546}. 
 Apart from the spectral signature, also the spatial
 extent of the \OI emission region is investigated.
 For both stars, we 
 have obtained spectra with the slit positioned at different position
 angles on the sky.}
{The \OI emission region of HD~100546 appears to be
 coinciding with the dust disk, its major axis located at
 150$\pm$11\deg~east of north. The SE part of the disk moves towards
 the observer, while the NW side is redshifted. 
 The \OI emission region rotates counterclockwise around the central
 star. For
 HD~97048, the position angle of the emission region is
 160$\pm$19\deg~east of north, which is the first determination of
 this angle in the literature. The southern parts of the disk are
 blueshifted, the northern side moves away from us. Our data support
 the idea that a gap is present at 
 10~AU in the disk of HD~100546. Such a gap is likely
 planet-induced. We estimate 
 the mass and orbital radius of this hypothetical companion
 responsible for this gap to be $20~M_\jup$ and 6.5~AU respectively.}
{Based on 
 temporal changes in the \OI line profile,
 we conclude that inhomogeneities are present in the \OI
 emission region of HD~100546. These ``clumps'' could be in
 resonance 
 with the suggested companion, orbiting the central star in about 11~yr. 
 If confirmed, these observations
 could point to the existence of an object straddling the line
 between giant planet and brown dwarf in a system as young
 as 10 million years.}

\keywords{circumstellar matter --- stars: pre-main-sequence ---
             planetary systems: protoplanetary disks --- planetary
             systems: formation --- stars: 
             individual: HD~97048 --- stars: individual: HD~100546}

\maketitle


\section{Introduction}

HD~97048 and HD~100546 are two of the most famous Herbig Ae/Be
stars. The shape of the mid-IR spectral energy distribution
(SED) suggests that their circumstellar disks have a flared geometry
\citep{dominik03}. For HD~100546, broad-band images
---sensitive to scattered light--- of
the disk structure have been obtained in the optical and near-IR
\citep{pantin00,augereau01,grady01}. Furthermore, the 
disk around HD~100546 has been spatially resolved at UV--to--mm
wavelengths in the continuum as well as in spectral features emanating
from the disk
\citep[e.g.][]{wilner03,liu03,vanboekel04,leinert04,grady05}.
The disk's appearance and position on the sky are hence relatively
well-established for this source: the inclination of the system is
$i$$\approx$50\deg\ and the position angle of the major axis is
approximately 150\deg\ east of north \citep[e.g.][]{liu03}.
The disk of HD~97048 on the other hand has not
been observed in scattered light, although the mid-IR (around
10~$\mu$m) observations of \citet{vanboekel04} show that the source is
resolved in both the continuum and the PAH bands.

Herbig Ae/Be stars are by definition emission line stars. In a
significant fraction of these objects, also forbidden transition
lines are observed. In a previous paper \citep[][hereafter
AVD05]{ackeoi}, we have studied the \OI lines at 6300 and
6363\r{A}. We have shown that for HD~97048 and HD~100546, the
spectral line profiles are in agreement with the theoretical profiles
emanating from excited neutral oxygen atoms in the surface of a
flared, rotating passive disk. We have demonstrated that these atoms
cannot 
be thermally excited, and suggested the excitation may be due to the
photodissociation of OH and H$_2$O molecules in the surface layers of
the disk by the UV radiation field of the central star.

Since both sources are relatively nearby and the emission region of
the \OI 6300\r{A} line is expected to be spatially extended, we have
tried to obtain spatially resolved high-resolution spectra of the
targets around this line. We present the data set and reduction method
in the following Section. The analysis of the data and the confrontation
with the model are described in Sects. \ref{analysis} and \ref{model}.
In Sect.~\ref{planet} we discuss new evidence for the presence of a
giant planet around HD~100546, based on the results of the analysis.
The mass and orbital radius estimates for this object ($M_\pl =
20~M_\jup$ and $R = 6.5$~AU respectively) agree strikingly well
with previous studies concerning the gap in HD~100546's disk
\citep{bouwman03,grady05} and the spiral-arm structure in the outer
disk \citep{quillen05}. The final conclusions are summarized in
Sect.~\ref{conclusions}.


\section{Data set and reduction method}

We have obtained a number of high-resolution (\res=80,000)
VLT/UVES\footnote{http://www.eso.org/instruments/uves/} 
spectra for HD~97048 and HD~100546, with the slit positioned at
different angles on the sky. 
For HD~100546 we have taken spectra
with the slit aligned with (PA~=~150\deg~east of north) and
perpendicular to (PA~=~60\deg~E of N) the known major axis of the system, as
well as at two intermediate PAs (15\deg~and 105\deg~E of N).
For HD~97048, no information on the disk's inclination
or major axis 
is present in the literature. Therefore, we have chosen random,
equally spaced PAs of 0\deg, 45\deg, 90\deg~and 135\deg~E of N.
The slit length is
12\arcsec, while its width is 0.3\arcsec. In
Table~\ref{dataset}, the observation log is presented.

UVES is an echelle spectrograph, which implies that the raw 2D CCD
frames contain one 
{\em spectral} direction (along the different orders, perpendicular to the
slit) and one {\em spatial} direction (perpendicular to the orders, along
the slit). The small width of the slit ensures the high
spectral resolution needed to spectrally resolve the
\OI line. In the spatial direction the spectra have a resolution of
0.182\arcsec/pixel, or equivalently $\sim$64 pixels over the
12\arcsec\ 
slit length. In a standard reduction scheme, the spectrum of the
target is retrieved 
by integrating the flux in the spatial direction. However, we intend to
use both the spectral {\em and} spatial information contained in the
echelle orders. Therefore, we have taken the 
pipeline-reduced 2D echelle frames as our starting point. Spectra obtained
with the same slit PA were averaged such that the peak of the
spatial profile in the continuum of both frames coincides. In the
following, we only consider the small part of the UVES CCD around the
6300\r{A} line in the averaged 2D spectrum at each slit PA.

\begin{table}[!bht]
\caption{The data set. For both stars, spectra were obtained at four
  slit PAs. 
 \label{dataset}} 
\begin{center}
\begin{tabular}{lcccc}
 \multicolumn{1}{c}{Object} &
 \multicolumn{1}{c}{Date} &
 \multicolumn{1}{c}{UT} &
 \multicolumn{1}{c}{Slit PA} &
 \multicolumn{1}{c}{Exp. time} \\
 \multicolumn{1}{c}{} &
 \multicolumn{1}{c}{dd/mm/yy} &
 \multicolumn{1}{c}{hh:mm:ss} &
 \multicolumn{1}{c}{[\deg~E of N]} &
 \multicolumn{1}{c}{[s]} \\
\hline
HD~97048  & 21/03/05 & 04:41:47 &   0  & 300 \\
          & 21/03/05 & 04:47:43 &   0  & 300 \\
          & 22/03/05 & 02:26:21 &   0  & 300 \\
          & 22/03/05 & 02:32:18 &   0  & 300 \\
          & 27/03/05 & 00:40:24 &   0  & 450 \\
          & 27/03/05 & 00:48:49 &   0  & 450 \\
          & 27/03/05 & 01:00:48 &  45  & 450 \\
          & 27/03/05 & 01:09:12 &  45  & 450 \\
          & 27/03/05 & 01:18:56 &  90  & 450 \\
          & 27/03/05 & 01:27:22 &  90  & 450 \\
          & 27/03/05 & 01:36:49 & 135  & 450 \\
          & 27/03/05 & 01:45:14 & 135  & 450 \\
\hline				       
HD~100546 & 27/03/05 & 01:58:59 & 150  & 105 \\
          & 27/03/05 & 02:01:40 & 150  & 105 \\
          & 27/03/05 & 02:05:25 &  15  & 105 \\
          & 27/03/05 & 02:08:06 &  15  & 105 \\
          & 27/03/05 & 02:12:03 & 105  & 105 \\
          & 27/03/05 & 02:14:44 & 105  & 105 \\
          & 27/03/05 & 02:18:36 &  60  & 105 \\
          & 27/03/05 & 02:21:16 &  60  & 105 \\
\end{tabular}
\end{center}
\end{table}

\OI emission of the Earth's atmosphere, the so-called airglow,
can easily be removed from the spectra. The scientific target only
fills a fraction of the slit, while the airglow feature is equally
strong in all spatial directions on the sky around the source and
hence fills the entire slit homogeneously. We have determined the  
spectral profile of the airglow feature by averaging 22 spatial rows
at the top and bottom edge of the echelle order (i.e. outside the
region of the slit which is dominated by the target). This average is
then subtracted from all (spatial) rows, which results in a 2D
airglow-corrected frame.

\begin{figure}[!thp]
\resizebox{0.5\textwidth}{!}{\rotatebox{0}{\includegraphics{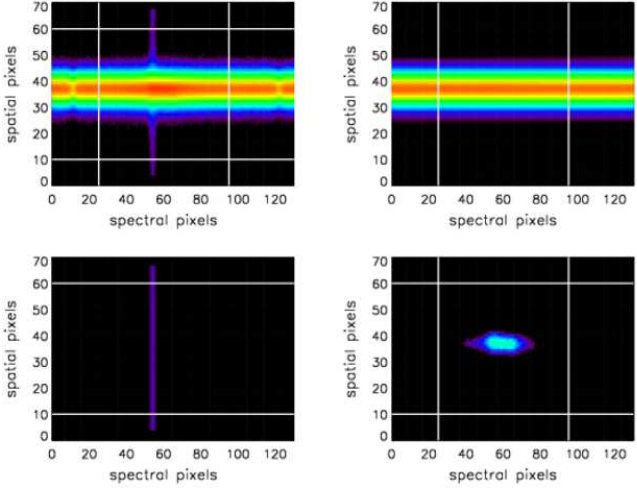}}} 
\caption{ The sky-subtracted 2D echelle frame ({\em top left panel})
  contains the 
  spatially (vertical) and spectrally (horizontal direction) dispersed
  spectrum of the target. The logarithmic color scale is the same in
  the four panels and represents the measured flux in each pixel. The
  bright horizontal band (order) is the 
  observed flux of the target, while the narrow vertical band, which
  fills the entire slit, is the airglow of Earth's atmosphere. The \OI
  feature is located approximately in the middle of the 2D frame.
  The two ``dips'' in intensity left and right of the feature are two
  telluric absorption lines. For all our spectra, these absorption
  lines are located outside the spectral range of \OI
  feature. Therefore there
  is no need to correct for this atmospheric effect.
  Based on the spatial rows ``above'' and
  ``below'' the echelle order (horizontal lines), the average spectral
  shape of the airglow feature can be determined. This average airglow
  feature ({\em bottom left panel}) is subtracted from the 2D science
  frame. The spatial profile of the continuum blue- and
  redward of the \OI feature (vertical lines) is used to compute the
  2D continuum ({\em top right panel}) by linearly interpolating in
  spectral direction for each spatial row. Finally, the
  continuum-subtracted, airglow-corrected 2D science frame, which only
  contains the \OI feature, is extracted ({\em bottom right
  panel}). \newline 
  The normalized 2D spectrum, which is not shown in this figure, is
  retrieved by dividing the airglow-corrected science frame by the 2D
  continuum. } 
\label{redproc.ps}
\end{figure}

Instead of integrating over the spatial direction and achieving the
``standard'' spectrum, we have determined a normalized spectrum
at each sampled spatial position. First we have
determined the continuum on the blue and red side of the line. We
average the continuum over 5 {\em spectral} pixels. Hence we obtain an
average 
{\em spatial} profile of the continuum blue- and redward of the \OI
line. To determine the interjacent continuum, a linear interpolation
in the spectral direction is done at each spatial pixel row. We end up
with a 2D continuum. The normalized spectrum of the source at each
sampled spatial position is then retrieved by dividing the flux in
each spatial row of the 2D echelle frame by the corresponding row of
the 2D continuum. This procedure is the straightforward 2D translation of
the ``standard'' normalizing procedure. Furthermore, a 2D
continuum-subtracted spectrum is computed by subtracting the 2D
continuum from the 2D airglow-corrected science frame.
In Fig.~\ref{redproc.ps} we
schematically summarize the airglow-removal and normalizing method
described above.

\section{Analysis \label{analysis}}

AVD05 have shown that the double-peaked profiles observed in the
\OI spectra of many Herbig Ae/Be stars emanate from excited oxygen
atoms in the surface layers of an inclined, rotating flared
disk. HD~97048 and HD~100546 are two relatively nearby
\citep[distances 103 and 180~pc for HD~100546 and HD~97048
respectively,][]{vandenancker98} Herbig stars that were shown to
display these double-peaked \OI emission lines in this study. If the
emission region is spatially 
extended, it is possible to extract spatial information on these
objects in the \OI line. When altering the
position angle of the slit on the sky, different spatially dependent
spectra are expected to be observed. If the slit is positioned over
the target along the major axis of the disk, the normalized spectrum
on one spatial side of the 
order will primarily show the blue peak of the profile, while the
spectrum on the other spatial side will predominantly display the
red peak. In the case of the slit oriented perpendicular to the
disk's major axis and centered on the star, no change in spectral
shape at different spatial positions is expected: the effect is wiped
out due to spatial integration in the direction perpendicular to the
slit. Intermediate PAs of
the slit on the sky lead to intermediate changes in spectral profiles.
No effect of slit PA is expected outside the emission line, because
the continuum consists of photospheric emission of the central star.

Apart from the normalized spectra at different spatial positions along
the slit, we have deduced a second quantity which may help to
interpret the observations. As a function of wavelength
(or equivalently radial velocity) we have determined the {\em spatial}
peak 
position of the echelle order in the {\em continuum-subtracted} 2D
frame. The 
quantity is deduced by fitting a Gaussian function to the spatial
profile in each spectral column \citep[this technique is called {\em
spectro-astrometry}, e.g.][and references therein]{takami04}. We
have compared these values to the 
peak position of the underlying continuum. If the slit is
positioned 
along the major axis, one expects to see the spatial peak position
shift from one side of the continuum peak position to the other when
running through the (spectral) line profile.
Together with the spatial peak position, the Gaussian fit delivers a
spatial full width at half maximum (FWHM). In the ---spatially
unresolved--- continuum this quantity is a measure for the atmospheric
seeing. In the \OI 6300\r{A} line the FWHM is potentially larger, if
the extent of the emission region is significant compared to the
atmospheric broadening. As can be derived
from Table~\ref{fwhms} this is not the case in our observations. The
FWHM ratios 
of the feature and the continuum are within the error bars all equal
to unity. We will come back to this point in Sect.~\ref{model}.

\begin{table}[!bht]
\caption{ The ratios of the measured FWHM in the feature and in the
  continuum, FWHM$_{feat}$/FWHM$_{cont}$. All ratios are within the
  error bars equal to unity. Based on our model assumptions
  (Sect.~\ref{model}), the expected maximum
  value for the ratio is 1.08 with the target at a distance of 103~pc
  and 1.03 at a distance of 180~pc. The slit PAs for the models
  indicated in this Table are relative to the modeled disk major
  axis (e.g., PA=0\deg~means slit and major axis aligned). 
 \label{fwhms}} 
\begin{center}
\begin{tabular}{lccc}
 \multicolumn{1}{c}{Object} &
 \multicolumn{1}{c}{Slit PA} &
 \multicolumn{1}{c}{FWHM$_{cont}$} &
 \multicolumn{1}{c}{FWHM$_{feat}$/FWHM$_{cont}$}\\
 \multicolumn{1}{c}{} &
 \multicolumn{1}{c}{[\deg~E of N]} &
 \multicolumn{1}{c}{[arcsec]} &
 \multicolumn{1}{c}{}\\
\hline
HD~97048  &   0 & 1.00 & 1.00 $\pm$ 0.04 \\
          &  45 & 0.93 & 1.00 $\pm$ 0.04 \\
          &  90 & 0.98 & 1.01 $\pm$ 0.04 \\
          & 135 & 1.13 & 1.00 $\pm$ 0.04 \\
HD~100546 &  15 & 1.32 & 1.01 $\pm$ 0.04 \\
          &  60 & 0.95 & 1.02 $\pm$ 0.03 \\
          & 105 & 0.98 & 1.01 $\pm$ 0.03 \\
          & 150 & 1.05 & 1.02 $\pm$ 0.03 \\
\hline
AVD05 180pc & 0 & 1.00 & 1.03 \\
            & 45& 1.00 & 1.02 \\
            & 90& 1.00 & 1.03 \\
            &135& 1.00 & 1.02 \\
AVD05 103pc & 0 & 1.00 & 1.08 \\
            & 45& 1.00 & 1.06 \\
            & 90& 1.00 & 1.08 \\
            &135& 1.00 & 1.06 \\
\end{tabular}
\end{center}
\end{table}

\begin{figure*}[!thp]
\center{\resizebox{0.95\textwidth}{!}{\rotatebox{90}{\includegraphics{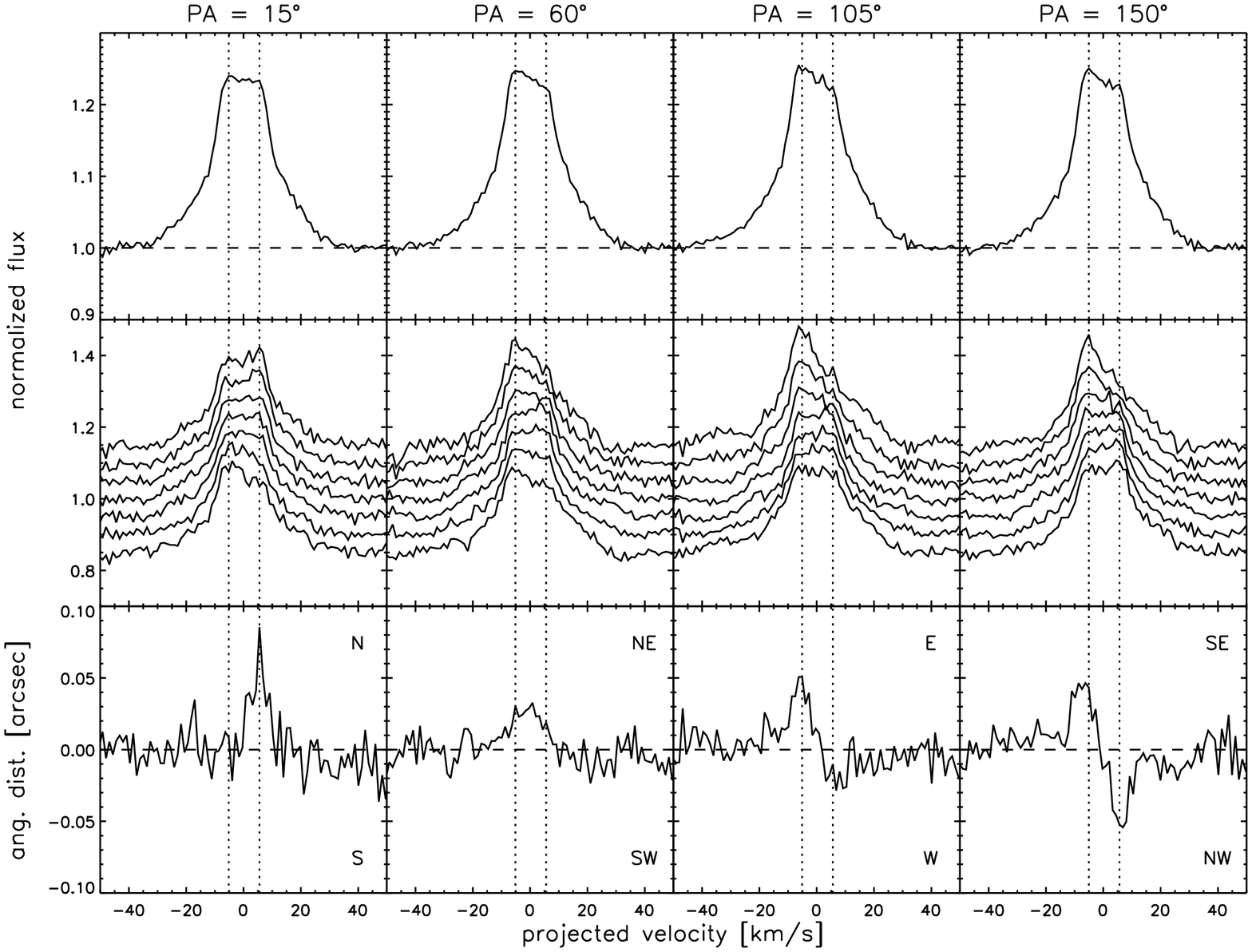}}}}
\caption{ 
  {\em Top panels:} The spatially integrated spectra of
  HD~100546 at different slit position angles (PA=15\deg, 60\deg,
  105\deg~and 150\deg~east of north). The normalized flux is plotted
  versus the (relative) radial velocity.  \newline
  {\em Middle panels:} The normalized spectra {\it at each spatial
  pixel row} around the central spatial position. The spectra are
  shifted for clarity. The formal spatial distance between two
  adjacent spectra is 0.182\arcsec, but atmospheric seeing blurs the
  spectra in spatial direction.
  \newline 
  {\em Lower panels:} The spatial peak position of the
  continuum-subtracted feature at each spectral row are plotted. The
  offset is relative to the spatial peak position of the underlying
  continuum. The scale is in arcsec. The curves are independent of the
  peak flux in each 
  spectral row and hence give a clear view on the spatial position of
  each (spectral) part of the line profile. In the top and bottom
  right of the figures, the positions on the sky ---in the direction of
  the slit PA--- are indicated. \newline
  The vertical dotted lines indicate the spectral position of the blue
  and red wings.}
\label{bigplotHD100546.ps}
\end{figure*}

\begin{figure*}[!thp]
\center{\resizebox{0.95\textwidth}{!}{\rotatebox{90}{\includegraphics{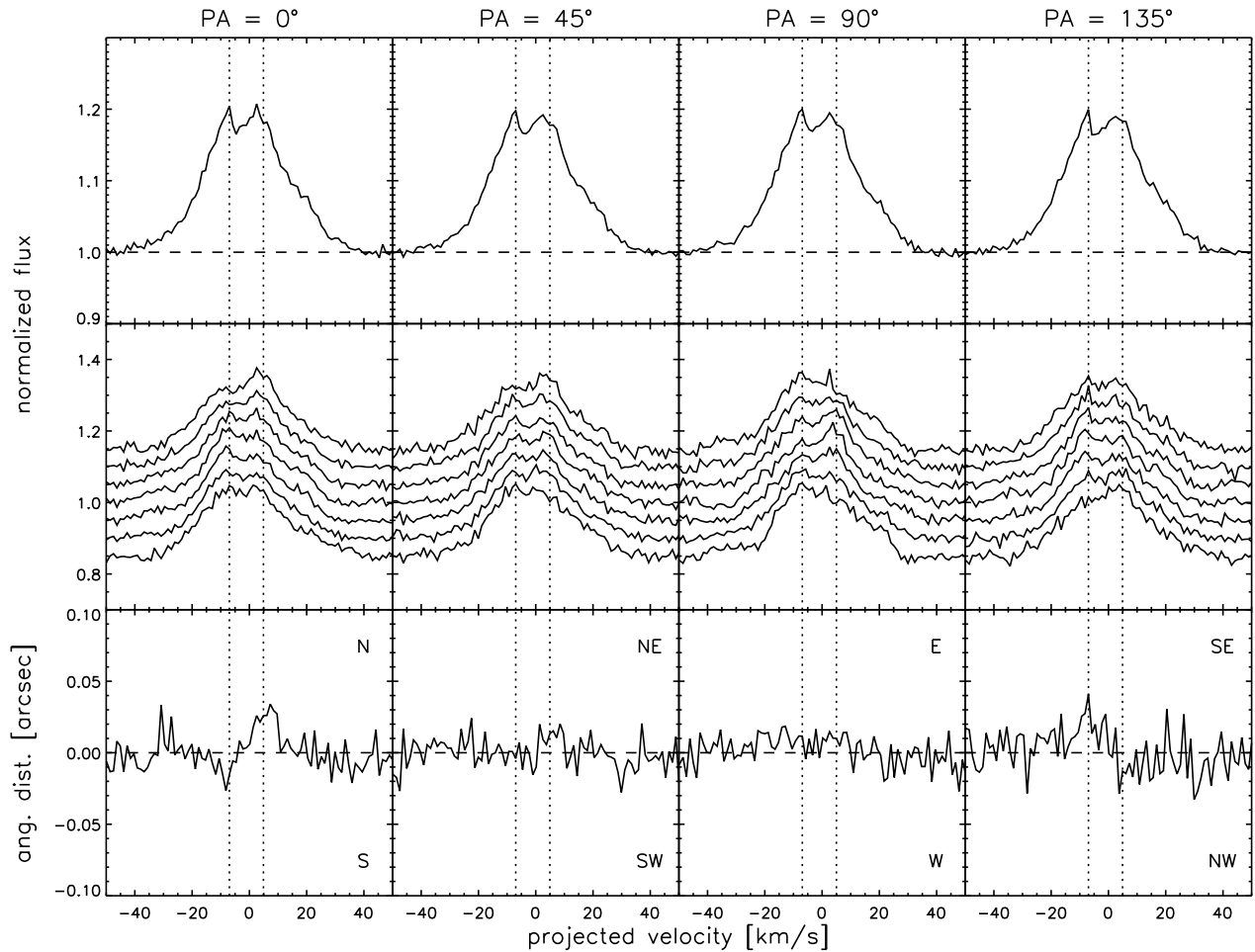}}}}
\caption{ Similar plot as Fig.~\ref{bigplotHD100546.ps} for HD~97048.} 
\label{bigplotHD97048.ps}
\end{figure*}

In Figs.~\ref{bigplotHD100546.ps} and \ref{bigplotHD97048.ps} we have
shown the (spatially integrated)
spectra of HD~100546 and HD~97048 at each slit PA (upper
panels). The small differences in spectral shape are likely due to the
small slit width (0.3\arcsec) and the extended targets, which implies
that only a part of the disk falls inside the slit. The 
second row of figures contains the normalized spectra at different
spatial rows around the central spatial position of the order. The
latter is the 
pixel row which contains the peak of the spatial profile of the
continuum. Note that, due to limited seeing, the spatial information
is blurred out over different spatial rows. Nonetheless, when moving
away from the central spatial row, the wings may become more
pronounced compared to the rest of the feature, if a small offset
of the blue or red emission region on the sky relative to the central
star is present. 
The bottom panels display the spatial peak position
of the continuum-subtracted \OI feature relative to the peak position
of the continuum. 
We have included this plot because
it is insensitive to asymmetry in the spectral line profile: for each
velocity, the determined central peak position is
independent of the total flux at that spectral position. These plots
give a more precise view on the spatial peak position of the blue and
red wing of the feature than the normalized spectra, in which the
relative peak flux of the wings influences the spectral shape at all
spatial rows due to atmospheric blurring.

The normalized spectra and spatial peak positions in
Figs.~\ref{bigplotHD100546.ps} and \ref{bigplotHD97048.ps} show the
signature of an emission region which is linked to the rotating
circumstellar disk. For HD~100546, the closest of both targets, this
fingerprint is clearest. With the slit positioned 150\deg~east of
north, i.e. along the disk major 
axis, the normalized spectra at the NW side of the star show the red
wing more pronounced than in the total spectrum, while the blue wing
is clearly dominant on the 
SE side. The same behaviour is seen in the shift in spatial peak
position in the feature relative to the continuum: the photocenter of
the red wing lies NW, and that of the blue wing SE of the
star. The same characteristics are seen in the spectrum at PA 105\deg,
although less strong. Due to the differential angle of 45\deg~between
the slit and major axis PA, the spatial extent of the \OI emission
region translates only partially into a spectral effect. At slit PA
60\deg, no difference in spatial peak position between the blue and
red wing is noted. The normalized spectra show no trend from SW to NE
either. However, the entire \OI feature seems to be shifted to the NE side of
the star. We will come back to this in Sect.~\ref{planet}. At PA
15\deg~east of north, the spatial profile is reversed: the slit
is turned 135\deg~on the sky and the north now appears on the top side
of the order. The peak position of the red wing is clearly on the N
side, confirming the conclusion derived from the spectra at slit PA
105 and 150\deg. The blue wing in the PA 15\deg~spectrum on the other
hand does not shift significantly to the S of the central star. In
Sect.~\ref{planet}, we discuss this anomaly. 

The distance to HD~97048 is estimated to be 180~pc
\citep{vandenancker98}. The extent of the \OI emission region is
therefore less clearly present in the data. Nonetheless the spatial
peak position in the spectra with PA=0\deg~suggests that the red peak
is located northward of the central star, while the blue peak lies to
the south. This behaviour is confirmed by the spatial peak positions
in the PA=135\deg~spectrum: the blue wing appears on the SE side, the
red wing on the NW side. The two spectra in between (at PA 45\deg~and
90\deg~east of north) do not display a clear spatial effect.

\begin{figure*}[!thp]
\center{\resizebox{0.95\textwidth}{!}{\rotatebox{90}{\includegraphics{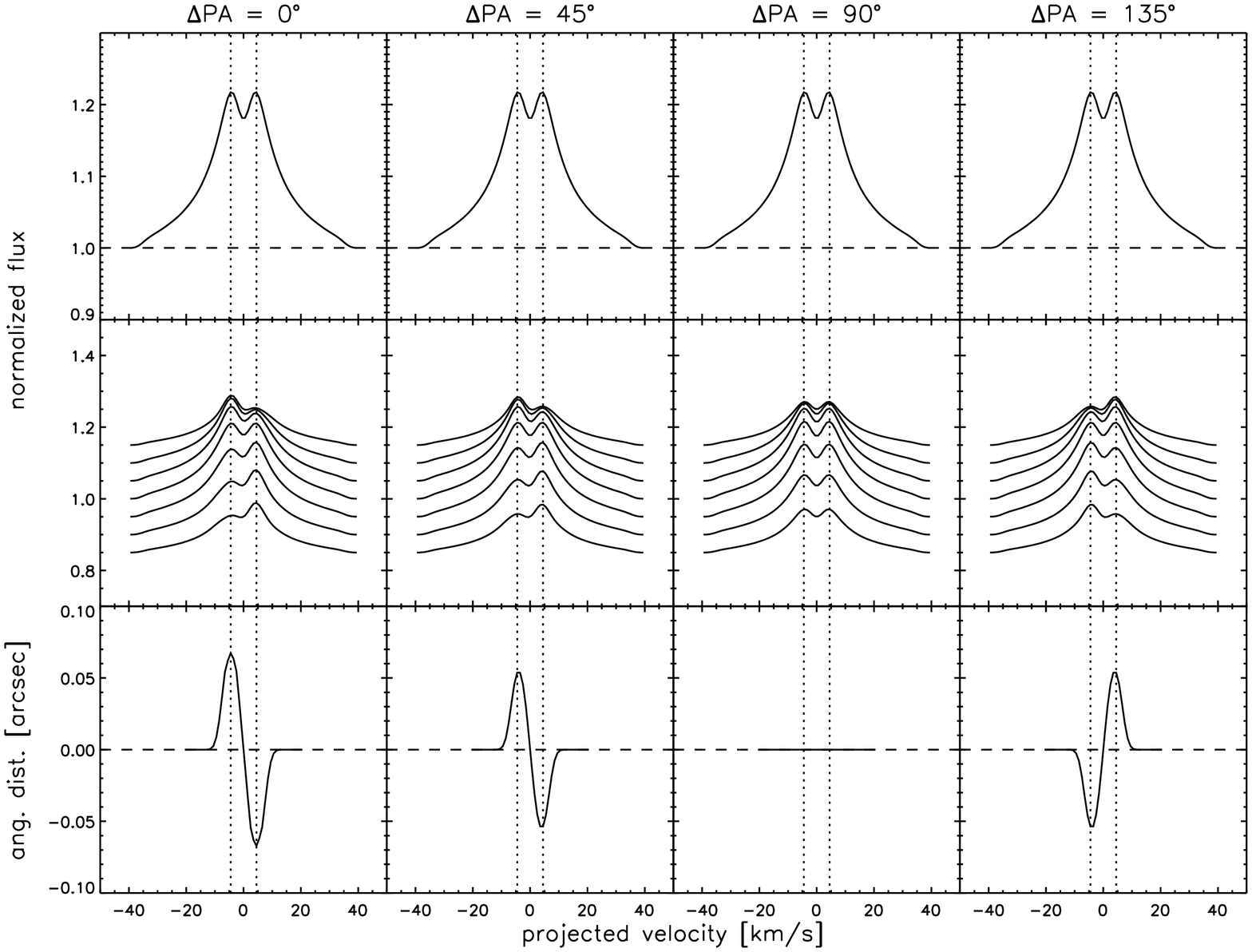}}}}
\caption{ Similar plot as Figs.~\ref{bigplotHD100546.ps} and
  \ref{bigplotHD97048.ps} for the
  model. If the PA of the slit is aligned with the PA of the major
  axis of the inclined disk ($\Delta$PA = 0\deg), the spatial
  effect is most pronounced. When slit and disk major axis are
  perpendicular ($\Delta$PA = 90\deg), no spatial effect is seen.
  The parameters of the model presented in this figure consist of
  the stellar 
  parameters of an A0V star, a surface-density power law $\Sigma
  \propto \mathrm{R}^{-2.5}$, an inclination $i = 45$\deg\ and a
  distance to the source of 180~pc (see AVD05 for
  details on the modeling). The line profile and continuum have been
  broadened with a Gaussian curve in spatial (FWHM~=~1\arcsec) and
  spectral direction (FWHM~=~3.7~\kms) to simulate the dispersion due
  to the atmosphere and the UVES spectrograph.} 
\label{bigplotD180.ps}
\end{figure*}


\section{Discussion: confronting the AVD05 model \label{model}}

\subsection{The model as starting point \label{modelstart}}

In AVD05 we have overplotted the spectral profiles of the \OI
lines of HD~100546 and HD~97048 with our model results (Fig.~26 in
that paper). In this
Section we go back to these models and confront not only their
spectral profile, but also their spatial extent with the UVES
observations. The parameters of
the model which we consider here include a stellar mass of
2.4~$M_\odot$, inclination $i$ = 45\deg\ and a surface-density profile
$\propto$~R$^{-2.5}$. 80\% of the \OI emission in our model emanates
from the region between 0.8 and 20~AU. For details on the modeling,
see AVD05. 

To simulate the effects of atmospheric blurring and
instrumental spectral broadening of UVES, we have convolved the
theoretical (spatially extended) emission line with Gaussian
profiles. In spatial direction,
we have taken a FWHM equal to 1\arcsec ---a typical value derived from
the UVES observations, see Table~\ref{fwhms}--- while in spectral
direction we take the instrumental width to be equal to the average
width of six (unresolved) telluric 
absorption lines near the \OI feature (3.7~\kms). The
results are presented in Fig.~\ref{bigplotD180.ps}. The four top-panel
spectra are identical, since we did not simulate the small slit
width, but just integrated over the entire disk. The middle panels
display the normalized spectra at each spectral row. The shape of the
spectrum changes depending on the relative position angle between slit
and disk major axis ($\Delta$PA) and the spatial distance to the
center of the target. The bottom panels show the theoretical peak
position throughout the feature.

Focusing on the
bottom row of panels, the following remarks can be made.
Qualitatively, the
effects seen in the modeled line profile agree with the observations,
indicating that the rotation of the disk indeed causes the
double-peaked \OI 6300\r{A} lines in HD~97048 and
HD~100546. Nevertheless, there is a anomaly in the {\em magnitude}
of the spatial effect: the shift of the peak positions in the modeled
profile is approximately three times larger than the observed
shift. The presented model is however not a fit to our data, but
an {\em ad hoc} solution. As we have shown in AVD05, this solution is
far from unique in explaining the observed (spectral)
profiles.
Moreover, the effects of atmospheric and instrumental
broadening have been mimicked by convolving the profile with Gaussian
curves. This is a first-order approximation. The real point spread
function may blur out the shift in spatial peak position even more or
at least differently.

\begin{figure}[!thp]
\center{\resizebox{0.5\textwidth}{!}{\rotatebox{90}{\includegraphics{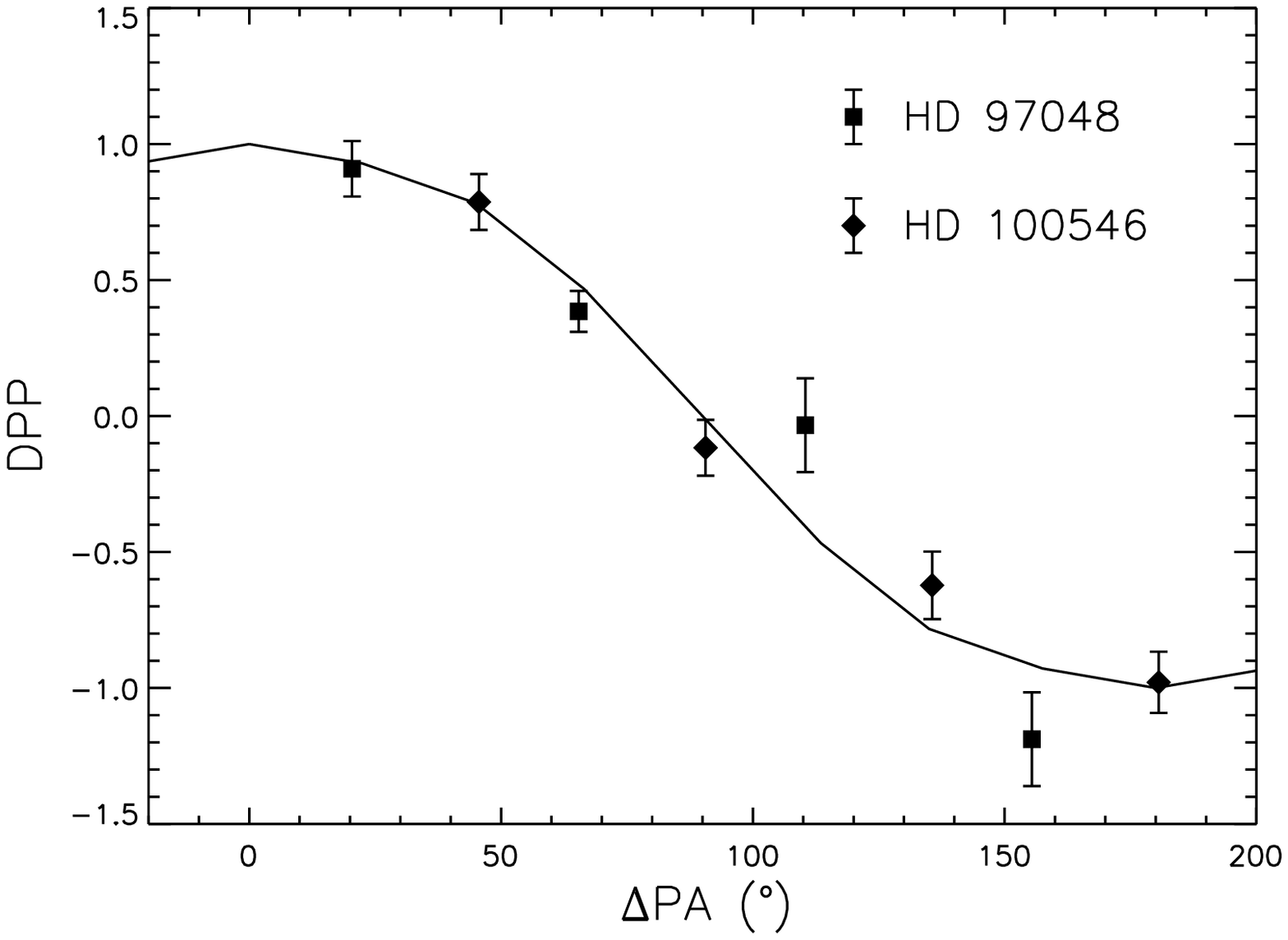}}}}
\caption{ The relative difference in peak position (DPP) between the
  blue and red wing for the model (solid line) and observations (squares
  and diamonds) versus the relative position angle ($\Delta$PA)
  of slit and disk major axis. The observed DPPs are
  normalized by dividing with the amplitude of the best-fit
  cosine. A relative shift of 0
  indicates that the spatial peaks of the blue and red
  wing of the feature are located at the same position.
  The best-fit ($\chi_\mathrm{red}^2$ = 1.9) PA of the disk
  major axis in HD~97048 is 
  160$\pm$19\deg~east of north. For
  HD~100546, the best-fit ($\chi_\mathrm{red}^2$ = 0.6) PA angle is
  150$\pm$11\deg~east of north, which
  coincides perfectly with the known PA of the disk.} 
\label{wobble.ps}
\end{figure}

To show that the shift in spatial peak position in both targets
is in qualitative agreement with the models, we have calculated the
{\em relative} difference in peak position (DPP) between the
red and blue wing. The spectral position of the wings is indicated
with the vertical lines in Figs.~\ref{bigplotHD100546.ps},
\ref{bigplotHD97048.ps} and 
\ref{bigplotD180.ps}. The relative peak position is computed for each
slit PA. For 
the models the values are normalized to unity by dividing them by
the maximum DPP (at $\Delta$PA = 0\deg). For HD~97048
and HD~100546, we have fitted the DPPs with a
cosine function $\mathrm{DPP} = A \cos(\mathrm{PA} - B)$ with $A$ and
$B$ fit parameters. The best-fit ($\chi_{\mathrm{red}} ^2$) 
amplitude $A$ is used to normalize the DPPs to unity. The best-fit PA
offset $B$ indicates the position angle of the major axis of the
disk. In Fig.~\ref{wobble.ps}, the modeled and observed DPPs 
are displayed. From the fit, we can derive that the PAs of the disk
major axis in HD~97048 and HD~100546 are respectively 160$\pm$19\deg\ and
150$\pm$11\deg~east of north\footnote{As a convention we have defined
the PA of the disk major axis {\em in 
the direction} of the blueshifted part of the disk.}. Note that the
---independently determined--- PA value for HD~100546's disk is
in perfect agreement with the known PA of this system
\citep[$\approx$150\deg~E of N, e.g.][]{liu03}.
We conclude that the S--SE parts of the disks around HD~97048 and
HD~100546 rotate toward us, while the N--NW side moves away from us.

Note that the spatial peak position of the feature in HD~100546 is not
symmetric. In the \OI line, the NE side of the order seems to be
brighter than the SW side, {\em pulling} the peak position ---which is
a measure for the photocenter--- in this direction. This asymmetry may
be due to the fact that the disk around HD~100546 is a flared disk:
the surface of the side of the disk closest to the observer is seen
under a more acute angle than the surface of the opposite side.
Therefore, the projected surface of the nearest side of the disk is
smaller than the projected surface of the opposite side. The modeling
of this effect is beyond the scope of the present paper. However, to
give an idea about its magnitude, we can make a simple
estimate. Assuming a typical, homogeneous flared disk with an opening
angle of 20\deg~and an inclination of 50\deg, the viewing angle
between the line of sight and the normal to the surface of the nearest
side of the disk is about 70\deg~ (i.e. almost ``edge-on''). The
viewing angle on the opposite side is approximately 30\deg. The
projected surface of the nearest side of the disk is under these
assumptions only $\sim$30\% of the projected surface of the other side
of the disk. As opposed to the spatial shift of the blue- and redshifted
side of the disk, which is maximal when slit and disk major axis are
aligned, this effect influences the spatial peak positions in 
a spectrum most when the slit and major axis are perpendicular. When
slit and major axis are aligned, the spatial 
integral ---perpendicular to the slit--- wipes out the signature of the
differences in projected surface. This is consistent with the
observations of HD~100546: at PA 15\deg\ and 60\deg, the effect may play a
role. This would indicate that the NE side of the disk is {\em further
away} from the observer than the SW side. If so, our knowledge
on the rotation of the \OI emission region can be further refined: not
only can we deduce that the blue wing is located on the SE side of the
disk, but also that the rotation occurs {\em counterclockwise} around the
central star. 

\begin{figure}[!thp]
\center{\resizebox{0.5\textwidth}{!}{\rotatebox{90}{\includegraphics{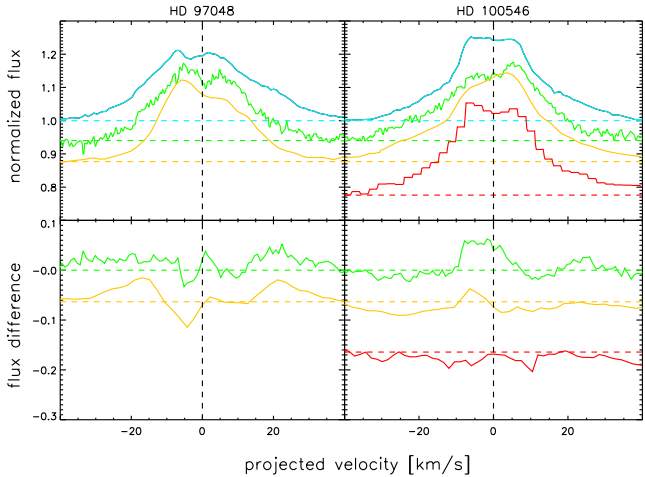}}}}
\caption{ Temporal changes in the spectra of HD~97048 and
  HD~100546. Top panels: the available normalized spectra, shifted
  according to their relative age (0.1 corresponds to 5\,yr). From
  top to bottom: 
  UVES, ESO~3.6m, FEROS and, for HD~100546 only, CAT. Bottom panels:
  the flux difference of the normalized UVES spectrum and the other
  normalized spectra, 
  again shifted according to relative age. These differential spectra
  are rebinned to the resolution of the lowest-resolution spectrum of
  the two. From top to bottom: $\mathrm{UVES}-\mathrm{ESO~3.6m}$,
  $\mathrm{UVES}-\mathrm{FEROS}$ and, for HD~100546 only,
  $\mathrm{UVES}-\mathrm{CAT}$.}
\label{tempchange.ps}
\end{figure}

A second effect which is not taken into account in our model is
possible inhomogeneity in the \OI emission region. Obviously,
deviations from homogeneity do occur: the spectral \OI feature is not
symmetric and temporal changes in the line profiles are observed. For
both targets, we have other medium-to-high resolution 6300\r{A}
spectra at our disposal (ESO~3.6m~CES and ESO~1.5m~FEROS, and CAT~CES
for HD~100546 only). These spectra are described in detail in AVD05.
In Fig.~\ref{tempchange.ps} the spectra are displayed. The plot
allow us to take a first glimpse at the {\em temporal} behaviour of
the \OI line in Herbig Ae/Be disks. The CAT spectrum (\res = 65,000)
is taken in 1994, the FEROS spectra 
(\res = 45,000) in 1999 and the ESO~3.6m spectra
(\res = 120,000) in 2002 (see AVD05 for the exact
dates). Assuming that emitting gaseous ``clumps'' survive in the disk
surface over multiple orbital periods, the temporal changes in the
spectral shape of the \OI line must display periodic behaviour, linked
to the disk's rotation. The respectively 3 and 4 spectra of HD~97048
and HD~100546 are hardly sufficient to describe potential periodicity
in the \OI 6300\r{A} line, but some conclusions can nevertheless be
drawn. First of all, the equivalent width ($EW$) of the \OI 6300\r{A}
line changes with time. For HD~97048, this change is small
($|EW|$=0.141, 0.135 and 0.160\r{A} in the FEROS, ESO~3.6m and UVES
spectrum) while it is larger for HD~100546 ($|EW|$=0.183, 0.180, 0.152
and 0.161\r{A} in the CAT, FEROS, ESO~3.6m and UVES spectrum). Also
the variations in spectral {\em shape} are larger in HD~100546. Due to
the short time base of the observations ---6 and 11\,yr for HD~97048
and HD~100546 respectively--- only the influence of fast orbiting
clumps can potentially be detected: for a 2.4~$M_\odot$ star, 6 and
11\,yr correspond to the orbital period at 4.5 and 7~AU
respectively. Inhomogeneities closer in orbit the star at least once
in this time span. Due to the fact that spectroscopic observations are only
sensitive to the {\em projected} velocity, non-homogeneous emission
from the high-velocity part of the disk influences the spectral shape
of the feature over the entire velocity range, and not only at high
projected velocities. For HD~100546, the double-peaked profile seems
to alternate between a blue-wing dominated (CAT, UVES) and a red-wing
dominated feature (FEROS, ESO~3.6m). We interpret this as additional
evidence for an \OI emission region which corotates with the
circumstellar disk. Furthermore, the presence of clumpy emission
provides an alternative explanation for the observed spatial peak
positions at different slit PAs. We come back to this point in
Sect.~\ref{planet}, when we discuss the possible origin of the
clumps.

\subsection{The observations as starting point \label{obsstart}}

The accuracy of our model can also be tested starting from the
observations. The spectral profile of the \OI 6300\r{A} line
contains information on the velocity field of the \OI emission
region. Assuming a Keplerian velocity field, a disk inclination and a
distance to the target, the spatial extent of the \OI emission region
can be calculated from the emission line. First 
we have averaged the blue- and red wing of the spectral profile in
order to construct a symmetric 
6300\r{A} line. The highest-velocity ({\vmax}) bin which contains \OI flux
(i.e. where the wing starts to stand out above the continuum) can
be linked to the innermost part of the \OI emission
region. The observed flux in this bin is attributed to a ring at
radius $R~=~GM_\star (\sin i/v_\mathrm{max})^2$. Only this inner ring
contributes to 
the flux in the {\vmax} bin, because the Keplerian velocity is smaller
at larger radii and because no \OI emission is detected at higher
velocities. We subtract the theoretical emission profile of this ring,
which covers all projected velocities from {$-$\vmax} to {$+$\vmax}, 
from the total spectral line profile. The spectral bin at {\vmax} of the
residual is now empty and we continue with the next
high-velocity bin. The method is repeated until the spectral profile
is completely dismantled down to $v$ = 0~\kms. In this way the observed
intensity-versus-radius 
profile can be reconstructed. For both HD~97048 and HD~100546 we have
taken $M_\star = 2.4~M_\odot$ and $i$ = 45\deg. In Fig.~\ref{ivr.ps}
we display the observed intensity-versus-radius curves, $I(R)$, deduced
from the spectral data of HD~97048 and 
HD~100546, and our modeled intensity distribution. We have multiplied
the curves with the surface $S(R)$ of the ring at radius R. In this way,
the relative contribution at each radius can readily be compared.
Since the modeled
flared disk is truncated at the dust destruction radius (0.8~AU in the
case of an A0V star), in our models the \OI emission region also
starts at this 
distance to the star. The outer limit is chosen to be 100~AU. Beyond
this limit, the modeled intensity drops quickly and is
insignificant compared to the \OI intensity closer to the star.
The observed $I(R) \times S(R)$ curve of HD~97048 has one prominent
peak. Under the current assumptions of stellar mass and inclination,
80\% of the total \OI emission emanates from the region between 0.9
and 12~AU. The curve of HD~100546 is qualitatively equal to that of
HD~97048 out to 5~AU, but shows a strong second
``bump'' further out, around 15~AU. 80\% of the total \OI emission
emanates from the region between 1.2 and 23~AU, with 25\% emanating
from the bump. The observations suggest that 
the \OI emission regions of HD~97048 and HD~100546 are more
concentrated around the central star than in our model. The
drop-off at large radii is significantly stronger than
predicted in both sources. The
inner edge of the model (0.8~AU) on the other hand is in agreement
with the observed intensity-versus-radius curve: no significant \OI
flux is detected at high velocities.

We have confronted the observational
intensity-versus-radius curves derived from the spectral line profiles
to the spatial extent deduced from the spatial information contained
in the UVES spectra. 
Assuming a stellar mass of 2.4~$M_\odot$, an
inclination of 45\deg~and a distance of 180~pc for HD~97048, the
spatial peak positions 
throughout the \OI line derived from the spatial and spectral
information agree. The maximum value of the ratio
FWHM$_{feat}$/FWHM$_{cont}$ derived from the observed
intensity-versus-radius curve is 1.01. This value is in agreement with
the values in Table~\ref{fwhms}: the spatial extent is not resolved
in the spatial FWHM of the feature.
Adopting the same stellar mass and inclination
for HD~100546 and a distance of 103~pc leads to a significant
difference in spatial peak 
position as derived from the spectrum-based intensity curve and the
spatial UVES information. The spatial peak position estimated from the
intensity curve is 1.5 times larger than the measured UVES peak
position. The 
ratio of the FWHM in the feature and in the continuum is 1.05. This is
on the verge of the spatial resolution of our UVES observations (see
Table~\ref{fwhms}). This factor may arise from an erroneous distance
to the source. The latter is needed to convert the radial scale to the
angular scale on the sky. The possibility that the distance to HD~100546
is underestimated by a factor of 1.5 seems however unlikely,
considering the small error bars ($\pm$7~pc) given by
\citet{vandenancker98}. In
order to reconcile both measurements, either the stellar mass of
HD~100546 has to be increased to 3.6~$M_\odot$ or the inclination
must be increased to 60\deg, or a combination of both proportional to
$M_\star \sin^2 i$.
When incorporating the 1.5 factor, the maximum
ratio of the FWHM in feature and 
continuum decreases to 1.03, undetectable in the current UVES
observations. The first peak in $I(R) \times S(R)$ would then shift
from 1.5--5~AU to 1--3~AU and the second \OI emission ``bump'' at
10--25~AU to 7--17~AU.  


\section{A massive planet orbiting HD~100546? \label{planet}}

\subsection{The presence of a gap reconfirmed}

\citet{bouwman03} have suggested that the mid-IR continuum emission in
the SED of HD~100546 points to the presence of a dust-poor cavity
around 10~AU. This 
cavity is believed to be planet-induced. Due to lack of dust at
10~AU, the 
far-side rim of the gap is directly irradiated by the central
star. The mid-plane temperature is therefore high and the vertical
scale height of the disk at the far side of the gap is more than a
factor of two that of a standard flared disk at the same radius. The
suggested gap and wall explain the relative lack of mid-IR continuum flux
shortward of 10~$\mu$m and the strong excess flux at longer
wavelengths. The presence of this gap has been confirmed by mid-IR
nulling interferometric measurements \citep{liu03} and {\em
Hubble Space Telescope} (HST) STIS observations \citep{grady05}. 
The latter authors have observed HD~100546 in the far-UV wavelength
range, obtaining long-slit spectra. They suggest that the gap has an
outer radius of 13~AU and is not centered on the star but rather on a
region 5~AU southeast of the star. They exclude that an
earlier-than-M5-type stellar companion is present within 2.5-10~AU of
HD~100546.

\begin{figure}
\center{\resizebox{0.5\textwidth}{!}{\rotatebox{90}{\includegraphics{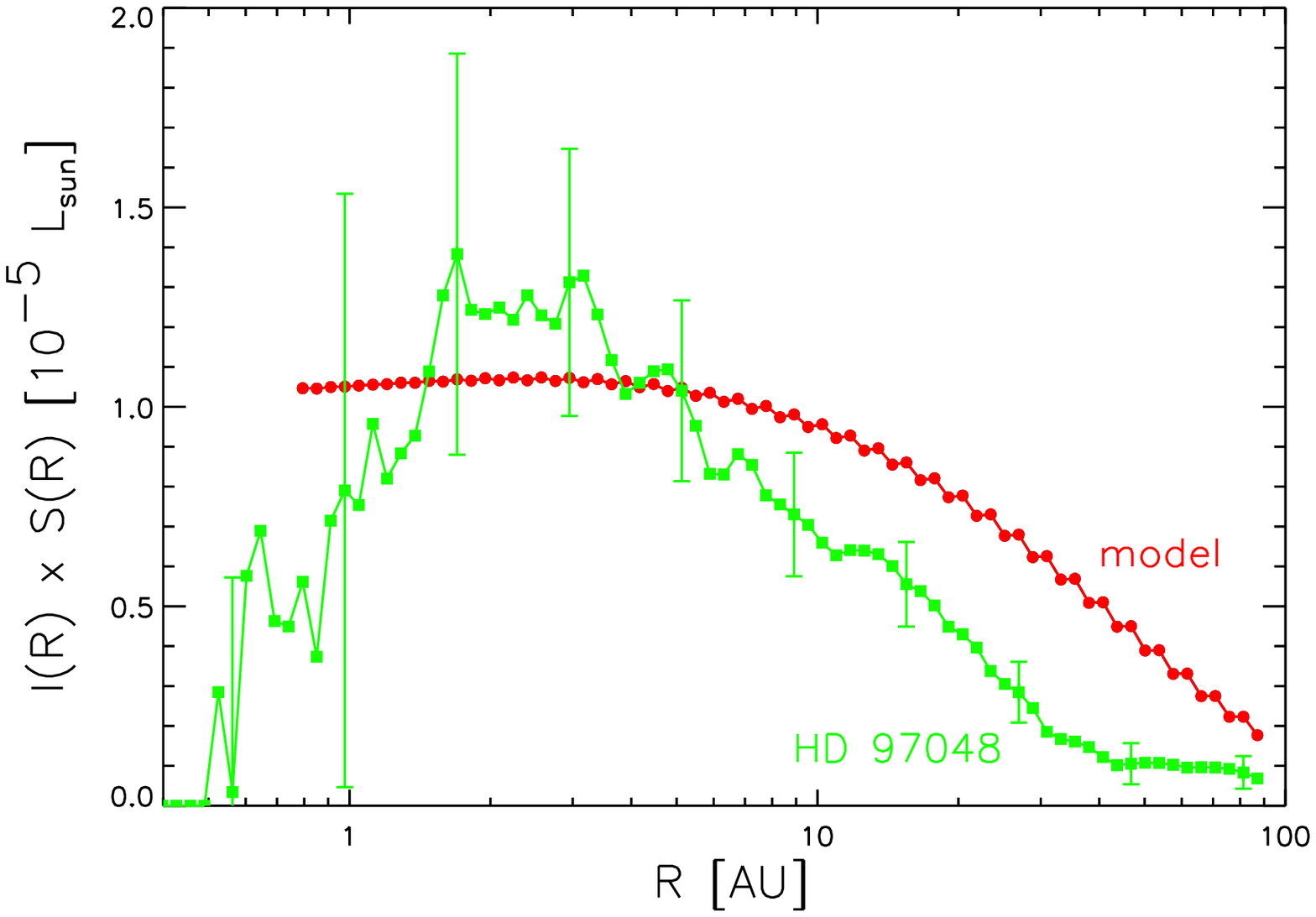}}}
\\
\resizebox{0.5\textwidth}{!}{\rotatebox{90}{\includegraphics{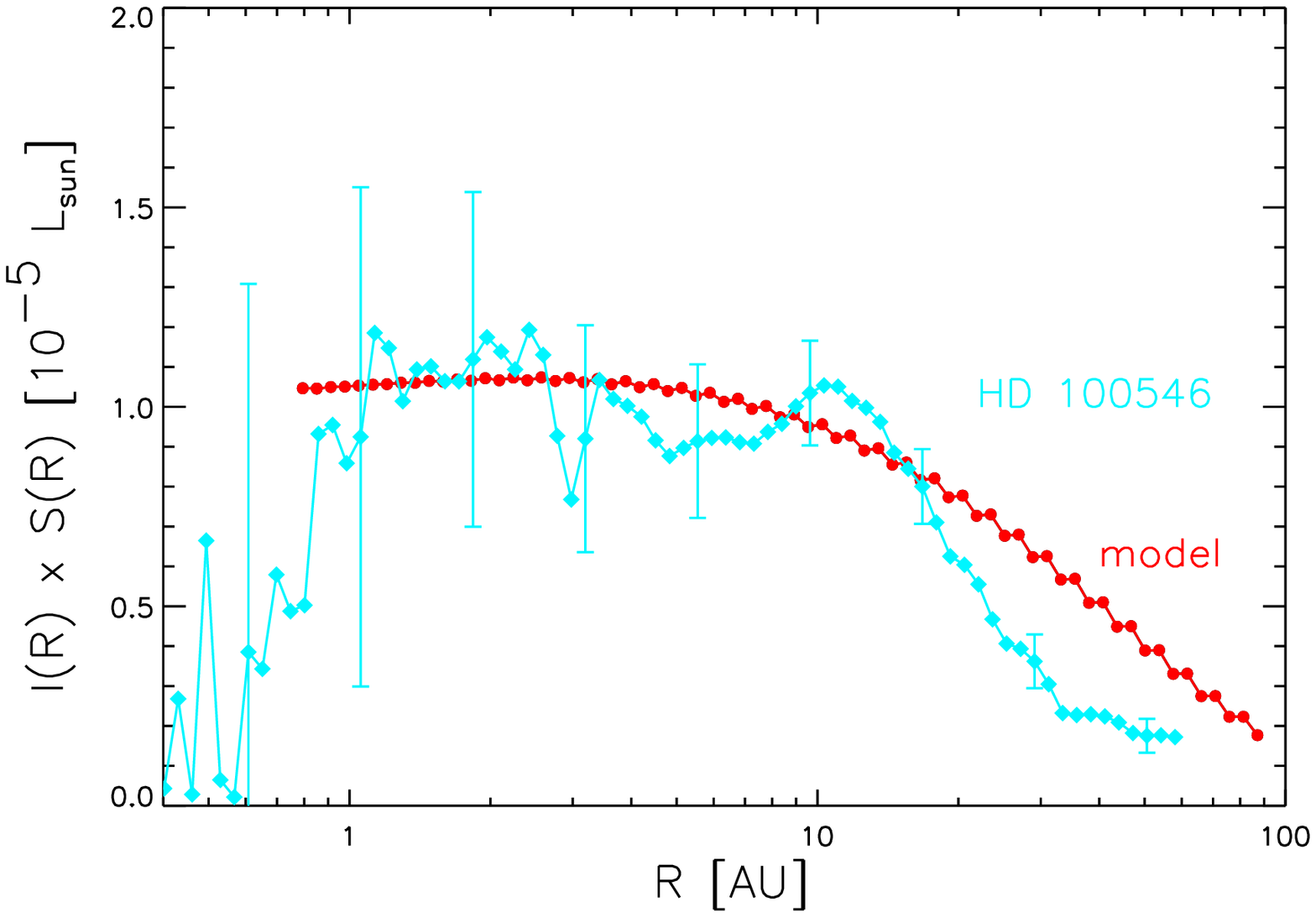}}}}
\caption{ The intensity-versus-radius curve $I(R)$ derived from the
  spectral shape of the \OI 6300\r{A} line in HD~97048 ({\em top}) and
  HD~100546 ({\em bottom}), and the model curve. The 
  intensities have been multiplied by the surface $S(R)$ of the ring at
  radius R. The fluxes of the inner and outer parts of the disk
  can be directly compared in this way: the total strength of the
  feature is the sum of the plotted values. The error bars have been
  deduced from the signal-to-noise ratio of the spectra. The model
  curve is only non-zero in the
  region between the dust destruction radius at 0.8~AU and the outer
  radius of 100~AU considered in the model. The
  observations suggest that the \OI emission regions in HD~97048 and
  HD~100546 are more compact than the model suggests. The peak of the
  $I(R) \times S(R)$ curve in HD~97048 is located between 1 and 5~AU.
  For HD~100546 a second peak can be discerned at 10--25~AU. See text
  for explanation.}
\label{ivr.ps}
\end{figure}

The intensity-versus-radius profile derived from the \OI feature in
HD~100546 (Fig.~\ref{ivr.ps}) may also be interpreted in this 
framework. 
The $I(R) \times S(R)$ curve in HD~100546 a local maximum at
short radii and declines furtherout up to $\sim$5~AU. The curve of
HD~97048 displays the same behaviour in this region, which suggests
that the inner parts of the \OI emission region are similar in both
sources.  
Further out however, the disk structure of HD~100546
appears to be significantly 
different than that of other Herbig 
Ae/Be stars such as HD~97048. Exactly where \citet{bouwman03} and
\citet{grady05} locate the outer edge of the gap, there is a
significant excess in \OI emission in HD~100546. Note moreover 
that the $I(R) \times S(R)$ curve of HD~100546 is continuous and has a
local minimum at 6--10~AU. The \OI emission region is suggested to be
located in the disk surface layer (AVD05). If an additional ``wall''
(apart from the inner rim at 0.8~AU) is present in the disk, the
irradiated disk surface is significantly larger than in the standard
flared-disk geometry. More \OI emission than in a standard flared disk
would be observed. Although the {\em dust} disk may be mostly cleared at
10~AU, the {\em gas} disk is very likely still present around this
radius \citep{bouwman03,grady05}. Therefore, the continuity of the \OI
intensity-versus-radius curve is not in disagreement with the possible
presence of a dust-free gap in the disk. The local minimum in the
intensity curve is located at the radius where the ``standard'' \OI
emission region ends and the ``additional'' wall emission begins.

The gap in the disk around HD~100546 is likely
planet-induced. The mass of the planet can in principle be deduced
from the size of the gap, the planet's orbital radius and the stellar
mass: according to \citet{artymowicz87} the diameter of the gap
cleared by a planet is equal to $4 \sqrt{3}~R_\mathrm{Roche}$, where
$R_\mathrm{Roche}$ is the radius of the planet's Roche
lobe. An additional constraint on the planetary mass can be obtained from
\citet{lin93}. The authors state that a gap can only be created if the
radius of the planet's Hill sphere \citep{hill1878} is 
larger than the vertical scale height of the disk. In their estimate
of the planet's mass, \citet{bouwman03} assume a relative scale height
$H/R = 0.045$ at 10~AU, based on the \citet{dullemond01} model for
\object{AB Aur}. In our \citet{chiang} flared-disk model for
HD~100546, $H/R$ is equal to $0.071$ at 10~AU
(AVD05). \citet{bouwman03} double the $H/R$ value arguing that the
vertical scale   
height of the wall at 10~AU must be twice as large as predicted by the
flared-disk models in order to explain
the mid-IR continuum fluxes. However, the suggested
wall develops due to the direct radiation of the central star. The
stellar photons can only reach the edge on the far side of the gap
{\em if a gap is already present}. In our opinion it is therefore not
necessary to assume a larger vertical scale height of the disk than in
the original flared-disk model. The lower limit to the companion's
mass is then $M_\pl > (H/R)^3~3~M_\star = 0.7$ or $2.7~M_\jup$
depending on the chosen flared-disk model.

\begin{figure}
\center{\resizebox{0.5\textwidth}{!}{\rotatebox{90}{\includegraphics{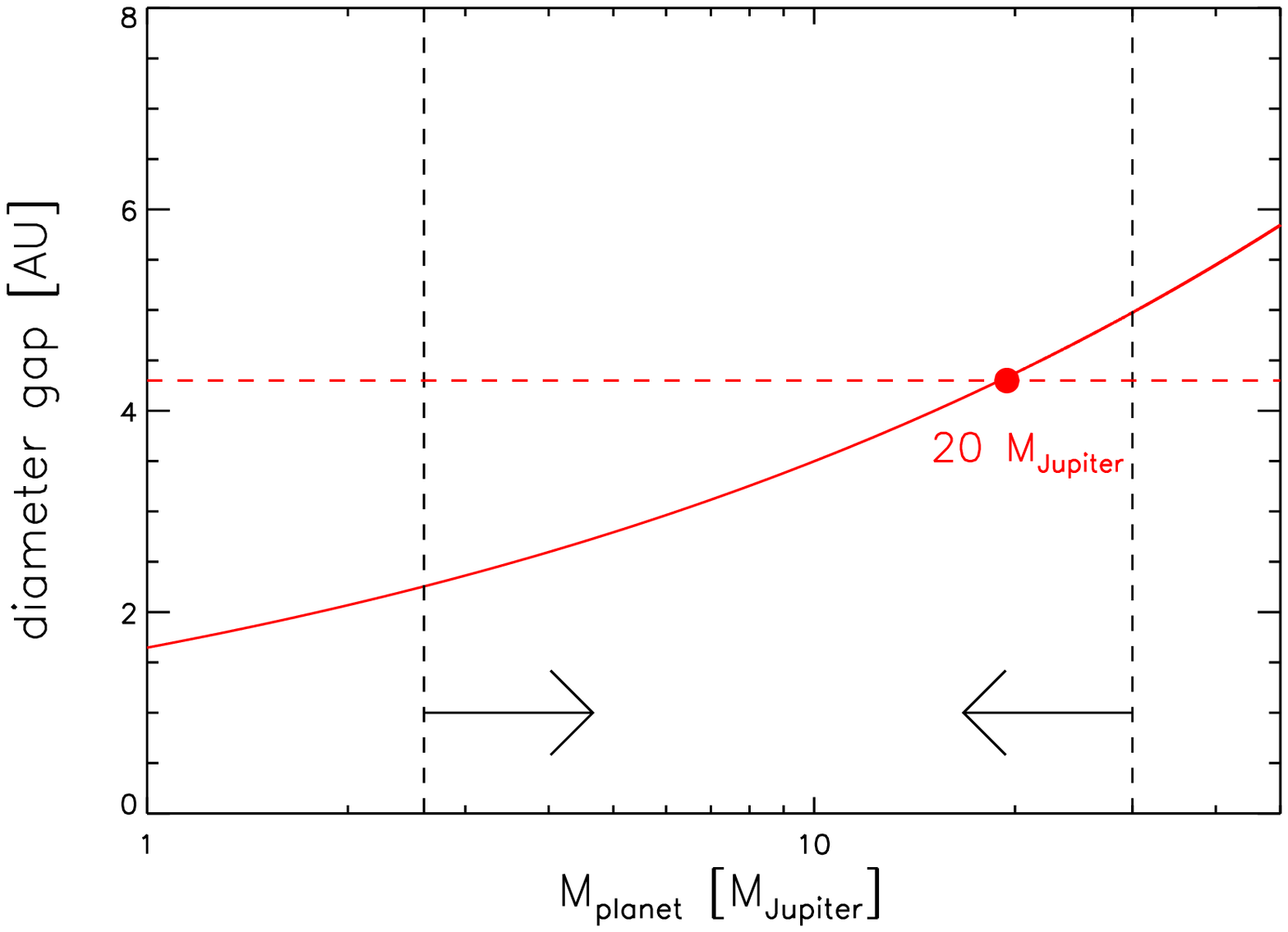}}}}
\caption{ The size of the planet-induced gap in the disk of HD~100546,
  $4 \sqrt{3}~R_\mathrm{Roche}$, as a function of the planet's mass
  \citep{artymowicz87,eggleton83}. The curve represents a companion
  at a 6.5~AU distance from the central star. The horizontal dashed
  line is an estimates for the diameter of the gap (4.3~AU). The
  curves intersect at a companion's mass of $20~M_\jup$. The left
  vertical dashed line and arrow indicate 
  the lower limit to the planet's mass, $2.7~M_\jup$,
  derived from the requirement that the planet's Hill sphere radius is
  larger than the local vertical scale height in the disk. The
  vertical line on the right side indicates the total disk mass in
  small dust grains, $30~M_\jup$, derived from the 1.3mm flux
  \citep[e.g.][]{ackesubmm}. \citet{grady05} exclude the presence of an
  earlier-than-M5 type companion. According to \citet{schmidtkaler},
  this corresponds to an upper limit on the mass of $\approx
  30~M_\jup$. We have indicated this upper limit with the second
  arrow.}
\label{Rocheradius.ps}
\end{figure}

In Fig.~\ref{Rocheradius.ps} the expected
diameter of the gap is shown as a function of the planet's mass. The
curve is based on the approximation for the Roche lobe radius
described by \citet{eggleton83}. The radius of the
planet's orbit is assumed to be the location of the local minimum in
the $I(R) \times S(R)$ curve of HD~100546, about 10~AU. We have
estimated the size of the gap by determining the FWHM around the local
minimum. The gap diameter we derive is 6.5~AU. Comparing the
theoretical and observed gap diameter, a mass of $20~M_\jup$
is deduced for the companion. An object with a 
mass exceeding $\approx 13~M_\jup$ lies beyond the deuterium-burning
threshold \citep{oppenheimer00} and thus the companion of HD~100546 is
a brown dwarf. Note that the spatial extent of the \OI
emission region in HD~100546 estimated from the $I(R) \times S(R)$ 
curve ---based on the {\em spectral} information in the UVES data---
differs by a factor 1.5 from the extent derived from the {\em spatial}
information in the UVES data. If we apply this factor, the distance
between planet and star 
becomes 6.5~AU and the gap diameter 4.3~AU. The estimated
mass remains $20~M_\jup$ however, since also $R_\mathrm{Roche}$
reduces by a factor of 1.5 in this case. An upper limit to the companion's
mass can be derived from the observational constraints set by
\citet{grady05}. They exclude the possibility of an earlier-than-M5 type
stellar companion. According to \citet{schmidtkaler}, an M4-type star
has a mass of $0.3~M_\odot$ or equivalently $30~M_\jup$. We have
indicated this upper limit in Fig.~\ref{Rocheradius.ps}.
As a reference, the total
disk mass derived from the 1.3mm flux of HD~100546 is also $30~M_\jup$
\citep{ackesubmm}. 

\subsection{Implications of the presence of a companion}

The rough estimates derived in the previous paragraph shed a new
light on other measurements of HD~100546. A massive companion disrupts
the original density distribution of the circumstellar disk, stirring
up the surrounding matter. Spots of over- and underdensities are
bound to form, their orbits synchronized with that of the companion. The
\OI emission region in the environment of the companion is therefore
expected to deviate from axisymmetry. The temporal changes in the line
profile of \OI 6300\r{A} line shown in Fig.~\ref{tempchange.ps} appear
to display variability with a typical
timescale of $\sim$11~yr. The suggested giant planet at 6.5~AU orbits the
central star 
in close to 11~yr. The ``clumps'' in the \OI emission region
of HD~100546 mentioned in Sect.~\ref{modelstart} may thus very well be in
{\em resonance} with the orbiting planet. \citet{grady05} derived from
the HST STIS long-slit spectra that the cavity is centered 5~AU
southeast of the star. The spectra were obtained on September 22, 2001
and June 23, 2002, i.e. approximately 6~months before and 3~months
after our ESO~3.6m CES spectrum of HD~100546. The latter spectrum
shows a \OI 6300\r{A} profile with a dominant red wing. Combining this
information with the newly obtained knowledge on the disk's rotation,
most of the \OI emission emanated from the northwest part of the
disk at that time. Assuming that the cavity is centered on the
giant planet's position hence leads to the conclusion that the gaseous
clump which causes the temporal changes in the \OI profile is located 
on {\em the opposite side} of the star as seen from the
companion. 

From the shape of the \OI 6300\r{A} line in the 2005 UVES spectrum of
HD~100546, the clump must mainly be located in the blueshifted part of
the 
disk. Nevertheless, the blue wing is not so dominant as the red wing
in the 2002 ESO~3.6m spectrum. This may indicate that the clump has
not reached its highest projected velocity or equivalently has not yet
arrived on the disk's major axis at PA=150\deg~east of north. This
image is reinforced by the shape of the spatial peak position through
the feature at different slit PAs (see bottom panels
Fig.~\ref{bigplotHD100546.ps}). The spatial peak position throughout
the profiles at PA=15\deg~and 60\deg~east of north appear to be
oriented more to the northeast/east than expected, especially in the
blue wing.
The deviation of the spatial peak positions from the modeled peak
positions at these PAs can be readily explained by the presence of a
\OI emitting clump between position angles 60\deg~and 105\deg~east of
north. 
The reasoning outlined above predicts that the planet is at present
(i.e. 2005) to the southwest/west of the star, between PA=240\deg~and 
285\deg~east of north. The spectral
shape of the \OI 6300\r{A} emission line thus provides a powerful tool
to prognosticate the companion's position on the sky.

The fact that the \OI emitting clump in the disk of HD~100546 is at
the time of the UVES observations (2005) to the northeast/east of the
star, and that the spectral profile was red-wing dominated in the
ESO~3.6m spectrum (2002) provides further knowledge on the rotation of
the \OI emission region and thus of the disk. In three years time,
i.e. approximately one quarter of the estimated orbital period, the
clump has moved from the redshifted region of the disk (i.e. the
north/northwest) to the north/northeast of the star. This indicates
that the emission region orbits the central star
counterclockwise. Note that we have already reached a similar
conclusion for the entire disk in
Sect.~\ref{modelstart}. However, the argumentation in that Section was
based on the assumption that the flaring of the disk's geometry causes
the deviation in the spatial peak positions of the UVES data. In the
present Section we have suggested that the signature in spatial peak
position can also be attributed to the presence of a clump in the \OI
emission. Even though both arguments may not be valid side by side,
the resulting conclusion (i.e. the disk's rotation occurs
counterclockwise) is the same. We believe that the presence of an
emitting clump provides the best explanation for the whole of
phenomena which give the \OI 6300\r{A}\ line its (temporal, spectral
and spatial) shape. The influence of the flaring of the disk is
probably less important.

The outer parts ($R \sim$ 300~AU) of the disk around HD~100546
display spiral-arm structure \citep{grady01}. In an effort to
explain this structure, \citet{quillen05} have modeled the effects of
a massive gravitationally bound object circling HD~100546. They
conclude that the mass of such an object needs to be of the order of
$20~M_\jup$ or larger to explain the observed impact on the disk
structure. They argue that the presence of such a massive companion at
$R <$300~AU is unlikely, because it would have been detected in the
HST images. They conclude that a recent fly-by of a nearby star is
preferentially the best explanation for the spiral arms,
although they cannot find a good candidate in the field. It is
striking however that our simple estimates for the companion's mass
and orbital radius agree with the boundary conditions set by 
\citet{quillen05}: the mass of the companion is sufficiently high to
cause the spiral 
pattern in the outer disk. Furthermore, the companion's orbit is so
close to the central star that the residuals of the stellar point
spread function (PSF) would easily cover up its presence in all
existing observations \citep[e.g.][]{augereau01}.

The presence of the $20~M_\jup$ companion described in this Section
leaves a signature in the radial velocities of the central star. The
mass center of the star-planet system lies at a distance of 0.12~AU
from the star. The amplitude of the stellar radial-velocity variations
caused by the massive planet hence is $0.33$~\kms~$\times \sin i =
0.25$~\kms~assuming $i$ = 50\deg. This amplitude is rather small,
considering the observational difficulties. The optical spectrum of
HD~100546 displays very few photospheric lines. Furthermore, these
absorption lines are broadened due to the stellar rotation
\citep[$v \sin i$ = 55~\kms,][]{ackelamboo}. Therefore such a companion
would easily go undetected in the existing spectroscopic data on
HD~100546.


With the current instrumentation available, it seems impossible to
directly observe the companion of HD~100546. The orbital radius of
6.5~AU corresponds to an angular separation of 65~mas at
103~pc. Space-based telescopes have diffraction limits which do reach
this kind of high-angular resolution. Adaptive Optics (AO) instruments on
ground-based 10~m telescopes are able to obtain this spatial
resolution. The main problem however is the large brightness contrast
between star and planet. Even in more favorable parts of the near-IR
spectrum (e.g. in a methane band), this brightness ratio is out of
reach for the present-day instruments. However, the companion should
be readily detectable with high-contrast AO instruments on future
ground-based Extremely Large Telescopes (ELTs).


\section{Conclusions and discussion \label{conclusions}}

In this paper we have provided further evidence that the \OI 6300\r{A}
emission in the disks around HD~97048 and HD~100546 emanates from the
disk surface, as was first suggested in AVD05. Thanks to the
combination of high spectral resolution and spatial
information on the \OI emission region contained in the UVES spectra,
we are able to determine the 
the blue- and redshifted part of the disk in the two targets. Although
the emission region was not resolved in the sense that the FWHM in the
feature is larger than the continuum FWHM, the observations clearly
show a variation in spatial peak position in the feature. The
dependence of that variation on the slit position angle agrees
qualitatively with our \OI 6300\r{A} emitting flared-disk model. 
Based on the observed intensity-versus-radius curve of HD~97048 and
HD~100546, we find that the \OI emission region in our model is more
extended than the emission region in the targets. 

The major axis of the well-studied circumstellar dust disk around
HD~100546 has a known position angle of 150\deg~east of north. We have
independently determined the same PA 
from our data. Note that two different diagnostics (the scattering
dust particles and the oxygen gas responsible for the \OI line) lead
to the same angle. This indicates that the \OI emission region is linked
to the geometrical distribution of the dust, in agreement with the
conclusions of AVD05. For HD~97048 we find a major axis PA of
160\deg~east of north, which is the first determination of this angle
in the literature.

Furthermore, we have obtained knowledge on the rotational
direction of both disks. The disk of HD~97048 is oriented
approximately north-south with 
the south part rotating towards us and the north part redshifted. The
disk of HD~100546 is positioned following a NW--SE line, with the SE
part blueshifted and the NW part moving away. 
For the latter object
we have shown that the rotation occurs counterclockwise around the central
star. The data for HD~97048 do not allow us to make a similar statement.

For HD~100546, we find evidence for the presence of a gap
in the circumstellar disk at about 10~AU, as was first suggested by
\citet{bouwman03} and later confirmed by \citet{grady05}. The radial
intensity distribution of the \OI 
emission region indicates the presence of excess emission at $R >
10$~AU in comparison to HD~97048. We suggest that this additional emission
emanates from the region on the far side of the gap in the dust disk,
where a wall has formed. The gap is likely
induced by a massive planet. Based on simple assumptions, we derive a
diameter for the gap of $\sim$4~AU and an orbital radius of 6.5~AU for
the planet. The mass of this object is estimated to be
$20~M_\jup$. The companion's orbital period ($\sim$11~yr) is
synchronized with the temporal changes observed in the \OI line
profile. We suggest 
that the deviations from axisymmetry in HD~100546's \OI emission
region are also planet-induced. The \OI line profile variations in
HD~97048, which displays no evidence for the presence of a massive
planet, are far less pronounced. The derived mass and orbital elements
of the companion of HD~100546 are in agreement with the conclusions of
\citet{quillen05}. The object cannot be directly detected in HST
images, and is massive enough to produce the large-scale spiral-arm
structure observed in the outer disk of HD~100546. The estimates
derived in the present paper can serve future efforts to detect the
companion orbiting HD~100546 by direct imaging with ground-based
Extremely Large Telescopes.

The \OI 6300\AA\ emission line in HD~100546 may prove to be a valuable
tool to probe the companion's orbital parameters. We will set up a
long-term monitoring campaign on the Swiss Euler 1.2m telescope in La
Silla (ESO) to study the variability of the line in more detail. The
results of this study can provide further refinements in our knowledge
of the giant planet's orbital period and hence distance to the
star. Monitoring the temporal changes of the \OI line may also lead to
a first estimate of the eccentricity of the planet's orbit.

The observations of HD~100546 mentioned in the present paper all point
to the presence of a massive object orbiting this target. Although the
estimated mass of the companion suggests that it is a brown dwarf, it
appears to have a formation history very alike that of a
planet. First, the object is expected to be close to HD~100546, which
is surrounded by a massive disk. Second, based on the 
mass estimate ($20~M_\jup$) the companion would appear in the ``brown
dwarf desert'': a lack of
binary companions in the mass range between $10~M_\jup$ and
late-M-type stars \citep[e.g.][and references therein]{mccarthy04}.
\citet{armitage02} propose that brown-dwarf companions form
contemporaneously with the primary. If the disk mass is sufficiently
high compared to the brown dwarf mass, disk-companion
interactions make the brown dwarf migrate inward. This occurs rather
rapidly ($\sim 1 \times 10^5$~yr for a massive disk). Given that the
observed mass of the disk around HD~100546 is at present still $\sim
0.3~M_\odot$ \citep[e.g.][]{ackesubmm}, the initial disk mass
should have been high enough to ensure rapid inward migration of the
companion. Furthermore, there does not seem to be an underabundance of
single stars throughout the brown dwarf mass range. Single stars form due
to self-gravitational contraction. Given the discrepancy between the
presence of single and long-lived companion brown dwarfs, is
unlikely that brown dwarfs in 
disk-dominated environments are formed following the same mechanism of
self-gravitation.  

On the other hand, a $20~M_\jup$ object in a massive disk
may have a similar formation history as a giant planet. Due to
gravitational instabilities in the disk, the object can gain mass and
grow \citep[e.g.][]{boss05}. If this mechanism is indeed at work in
massive protoplanetary 
disks, it appears logical that it favors low-mass, planet-like objects
over somewhat higher-mass brown dwarfs. Together with the fast removal
of initially formed brown dwarf companions, this would explain the brown
dwarf desert which is observed in multiple systems.
The HD~100546 system is at least 10 million years old
\citep{vandenancker98}. The presence of a massive object is therefore
more logical under the assumption of a formation which occured in the
disk (long) after the formation of the central star.
We conclude that the object orbiting HD~100546 is a rare specimen of a
brown dwarf companion with a formation history which is similar to
that of a regular planet.

The disk of HD~100546 is peculiar and extraordinary in many
ways. The longevity of the disk is striking in comparison with other
pre-main-sequence stars. Most disks dissipate within a few $10^5$~yr,
while the disk around HD~100546 is still clearly present. 
Also the presence of relatively strong accretion is exceptional
in such an old system \citep[Balmer line emission and accretion events
detected in the UV, e.g.][]{deleuil04,grady05}. 
The presence of a planet in the inner parts of the disk would 
cause a flow of gas which is ``funneled'' towards the central star.
Furthermore, the
solid-state features observed in the near-to-mid-IR suggest a highly
processed dust component, quite similar to the spectra of comets in
our solar system \citep[see e.g. the {\em Infrared Space Observatory}
data described by][]{malfait98b}. The degree of crystallinity
of the dust disk makes HD~100546 an outlier in the group of Herbig
Ae/Be stars \citep{ackeiso}. The grain size distribution deduced from
the 
(sub-)mm slope of the spectral energy distribution indicates the
majority of the emitting dust grains have small sizes ($\ll$1~mm),
while the dust particles in a lot of younger systems underwent
substantial grain growth \citep[$>$1~mm,][]{ackesubmm}. Now that it
becomes more 
and more clear that a $20~M_\jup$ companion is present, it is likely
that the characteristics which make HD~100546 special all find their
roots in the interactions between the disk and this companion.
Such interactions clearly maintain the disk. Based on the presence of
small and 
highly crystalline dust particles, we suggest however that the
observed disk is not pristine, but rather more evolved. The
interactions of the companion and the larger grains and planetesimals
in the disk ---which are expected to have formed in an earlier
stage--- could tear apart the latter. The small grains which emanate
from such destructive interferences then replenish the small grain
population in the disk. The occurance of events like these can
provide a satisfying solution to the unusual longevity of the
disk in some young stellar systems.

A final pecularity which is worth mentioning in this context concerns
the unusual photospheric abundances of HD~100546. HD~100546 appears
to be a so-called {\em $\lambda$~Boo star}: the photosphere
displays a deficit in metals (Mg, Si, Fe and Cr), while nitrogen and
oxygen have approximately solar abundances
\citep{ackelamboo}. HD~100546 is the only Herbig star in the sample of
\citeauthor{ackelamboo} which clearly displays this behaviour. {\em
  Selective accretion} of metal-depleted gas is commonly inferred to
explain this selective depletion pattern. The metals condense
into dust particles at much higher temperatures than the CNO
elements. The radiation pressure on dust grains prevents the latter to
fall in, while the gas-state elements are accreted. The gas and dust
should however be decoupled in order to prevent the dust to be dragged
along onto the star. The presence of a 
giant planet in the inner parts of the disk around
HD~100546 could create the specific conditions needed to produce the
selective accretion and consequent photospheric depletion
pattern. If most metals are locked inside dust particles
(or even planetesimals), the gas that is ``funneled'' by the planet
from the disk onto the central star is metal poor. Intriguingly, a
deficit of metals in the stellar photosphere is opposite to what is
typical of most giant planet systems found to date, in which the stars
are primarily metal rich.

Many observational characteristics of HD~100546 can be explained by
the presence of a giant, nearby planet. If the argumentation in the
present paper is confirmed by future observations, the companion of
HD~100546 is the youngest exo-planet discovered until today. The
constraints deduced from a study of such a young planet+disk system
will prove to be extremely valuable in both the modeling of
disk-planet interactions, as well as in improving our 
comprehension of disk evolution and planet formation.


\acknowledgements{The authors want to thank the ESO User Support
  Department, in particular F.~Primas and B.~Wolff, for their accurate
  and quick response to our questions concerning the UVES
  data. We also thank J.~Bouwman, C.~Dullemond, C.~Waelkens and
  L.~Waters for useful discussions. The referee, dr. S.~Edwards, is
  acknowledged for her constructive comments which helped to improve
  the clarity of the paper.}

\bibliographystyle{aa}
\bibliography{4330art.bib}

\end{document}